\documentclass[%
 reprint,
 amsmath,amssymb,
 aps,
]{revtex4-2}

\usepackage{graphicx}% Include figure files
\usepackage{dcolumn}% Align table columns on decimal point
\usepackage{bm}% bold math

\usepackage[utf8]{inputenc} % allow utf-8 input
\usepackage[T1]{fontenc}    % use 8-bit T1 fonts
\usepackage{hyperref}       % hyperlinks
\usepackage{url}            % simple URL typesetting
\usepackage{booktabs}       % professional-quality tables
\usepackage{amsfonts}       % blackboard math symbols
\usepackage{nicefrac}       % compact symbols for 1/2, etc.
\usepackage{microtype}      % microtypography
\usepackage{lipsum}
\usepackage{fancyhdr}       % header
\usepackage{graphicx}       % graphics
\graphicspath{{media/}}     % organize your images and other figures under media/ folder

\usepackage{amsmath}
\usepackage{graphicx}
\usepackage{dcolumn}
\usepackage{bm}
\usepackage{epsfig}
\usepackage{amssymb, latexsym}
\usepackage{graphicx}
\graphicspath{{./}{figures/}}
\usepackage{color}
\usepackage{xcolor} 
\usepackage{mathtools} 
\usepackage{orcidlink}

\usepackage{booktabs, multirow} % for borders and merged ranges
\usepackage{soul}% for underlines

\begin{document}

\title{Running gravitational constant induced dark energy as a solution to $\sigma_8$ tension} 

\author{Tilek Zhumabek}
 \email{tilek.zhumabek@nu.edu.kz}
 \affiliation{Department of Physics, School of Sciences and Humanities, Nazarbayev University, Astana 010000, Kazakhstan.}%Lines break automatically or can be forced with \\
 \affiliation{Energetic Cosmos Laboratory, Nazarbayev University, Astana 010000, Kazakhstan}

\author{Azamat Mukhamediya}%
 \email{azamat.mukhamediya@nu.edu.kz}
\affiliation{%
 Department of Electrical and Computer Engineering, School of Engineering and Digital Sciences, Nazarbayev University, 
Astana 010000, Kazakhstan
 }%

\author{Hrishikesh Chakrabarty}
 \email{hrishikesh.chakrabarty@nu.edu.kz}
\affiliation{
 Department of Physics, School of Sciences and Humanities, Nazarbayev University, 
Astana 010000, Kazakhstan % with \\
}%

\author{Daniele Malafarina}
 \email{daniele.malafarina@nu.edu.kz}
\affiliation{%
 Department of Physics, School of Sciences and Humanities, Nazarbayev University, 
Astana 010000, Kazakhstan 
}%

\date{\today}% It is always \today, today,

\begin{abstract} 
We consider a modified gravity model
%, dubbed GCDM, 
with a running gravitational constant coupled to a varying dark energy fluid and test its imprint on the growth of structure in the universe. 
Using Redshift Space Distortion (RSD) measurement results, we show a tension at the $3 \sigma$ level between the best fit $\Lambda$CDM and the corresponding
%Planck18/$\Lambda$CDM 
parameters obtained from the Planck data. 
%($w_0 = -1$, $\Omega_\mathrm{m} = 0.31$, $\sigma_8 = 0.81$).
Unlike many modified gravity-based solutions that overlook scale dependence and model-specific background evolution, we study this problem in the broadest possible context by incorporating both factors into our investigation. 
%By employing the Effective Fluid approach, the proposed model can be `mapped' to general relativity with a time-varying dark energy fluid to obtain background quantities. 
We performed a full perturbation analysis to demonstrate a scale dependence in the growth equation.   
Fixing the scale to $k = 0.1 h$ Mpc$^{-1}$ and introducing a phenomenological functional form for the varying Newton coupling $G$ with only one free parameter, we conduct a likelihood analysis of the RSD selected data. 
The analysis reveals that the model can bring the tension level within $1 \sigma$ while maintaining the deviation of $G$ from Newton's gravitational constant at the fifth order.    

\end{abstract}

\maketitle

\section{Introduction}
The standard model of cosmology, namely the $\Lambda$CDM model, provides an excellent description of the expansion history of the universe that ties together the density fluctuations observed in the cosmic microwave background (CMB) with the current observations of large scale structures (LSS) \citep{Planck:2018vyg,BOSS:2014hhw,BOSS:2014hwf,Baxter:2016ziy}. Over the last three decades, the measurements of the cosmological parameters have been continuously improving and at present we are in the era of precision cosmology. The parameters of the $\Lambda$CDM model have been measured with remarkable accuracy by the Planck mission in recent times and we use these values to define the standard model as Planck18/$\Lambda$CDM \citep{Planck:2018vyg}. However, the model seems incomplete as the nature of its major constituents dark energy (DE, which appears to be well described by the cosmological constant $\Lambda$) and cold dark matter (CDM) are still unknown. At the same time, the model requires an ad-hoc period of accelerated expansion in the early universe in order to have a nearly scale invariant initial condition \citep{Starobinsky:1980te,Guth:1980zm,Linde:1981mu}. 

To make matters worse, the increased precision of current observations and new analysis techniques have brought to light several tensions between values of different parameters among different observables and experiments. For example, some independent cosmological observations in the redshift range $z\lesssim 0.6$ are in disagreement with Planck18/$\Lambda$CDM. Two of such major tensions are one in the local measurement of the Hubble parameter $H_0$ (the so-called Hubble tension), and another in the estimates of the amplitudes of the matter power spectrum on the scales of $8 h^{-1}$ Mpc (often referred to as $\sigma_8$ tension). Both appear to be in conflict with the corresponding values obtained from Planck18/$\Lambda$CDM. On the other hand, recently it has also been shown that $\Omega_m$ increases with effective redshift in the late-time data \citep{Colgain:2022nlb,Colgain:2024ksa}, which may pose a challenge to the Planck18/$\Lambda$CDM model. Although the tensions might arise as a result of systematic errors, their persistence after several years of data analysis hints towards a possible breakdown of the standard scenario and the eventual requirement of new physics (see e.g. \citep{Abdalla:2022yfr} for a recent review).  

%{\bf (needs rephrasing and improvement)} 
In the present article, we shall attempt to address the so-called $\sigma_8$ tension. The parameter $\sigma_8$ describes the variance of the linear matter perturbations on scales of the order of $8 h^{-1}$ Mpc and it can be estimated from several observations, most notably, the weak lensing correlation function obtained by the CFHTLenS collaboration \citep{Heymans:2013fya}, the galaxy cluster count \citep{Planck:2013lkt} and the Redshift Space Distortion (RSD) measurements \citep{Macaulay:2013swa,BOSS:2016wmc,Nunes:2021ipq}. At present there exists a $2\sigma$ tension between the constraints of these local measurements on the matter density $\Omega_m$ and the amplitude $\sigma_8$ of matter fluctuation, and those from the Planck mission \citep{Battye:2014qga,MacCrann:2014wfa}. This tension can be reconciled by modifications to the standard scenario and in the literature, there are several suggestions to alleviate the tension such as different clustering of dark matter at smaller and larger scales \cite{Kunz:2015oqa}, a hot dark matter component \citep{Hamann:2013iba}, dark energy-dark matter interaction \citep{Lucca:2021dxo,Poulin:2022sgp,DiValentino:2019jae,Asghari:2019qld,BeltranJimenez:2021wbq,Jimenez:2024lmm} and modifications of general relativity (GR) \citep{G_parametrized, saridakis2023solving, Heisenberg:2022gqk}. However, many of these solutions are analytically complex, leading to the use of Planck18/$\Lambda$CDM background, or $H(a)$ for simplicity, which can induce undesired bias in the cosmological parameters \citep{Cardona_2021}. On the other hand, there exist some proposed models to alleviate the $H_0$ tension, for example with an early dark energy component, which have consequences for the $\sigma_8$ tension (see e.g. \citep{Poulin:2018cxd,Niedermann:2023ssr,Vagnozzi:2021gjh,Vagnozzi:2023nrq}). It would be preferable that any proposal to alleviate one tension would not make the other worse. 
In this context, examining simpler, less intrusive modifications becomes essential. 

In this article, we consider a simple modification of GR \citep{Markov:1984nw} (which we shall call the Markov-Mukhanov or MM model), which introduces in the Einstein-Hilbert action a multiplicative coupling of gravity and geometry through a scalar function of the energy density $\xi (\rho)$. 
This modification in the action leads to a running Newton's constant $G (\rho)$ and a corresponding induced dark energy component $\Lambda (\rho)$, both of which are related to the coupling function. 
In principle, one is free to choose one of the three functions $\xi$, $G$ or $\Lambda$ and obtain the others from the model.

Here we consider a model-independent parametrization of Newton's constant and obtain the variable $\Lambda$ from the MM prescription. We then perform a numerical analysis to constrain the parametrization of $G$ in our model from late-time structure formation data. We show that the Markov-Mukhanov model can help reduce the $\sigma_8$ tension, thereby shedding light on the proposed running and coupled nature of Newton's constant and dark energy. 
%of the underlying cosmological model. 
Our findings hint towards a slightly more repulsive force (or slight increase in the running dark energy density) in the early times of matter domination which agrees with the results reported in \citep{Nesseris_2022}. However, the difference between the present analysis and \citep{Nesseris_2022} is that, we do not use the Planck18 baseline and let $H(a)$ vary according to the specific dynamics of the MM model.

The article is organized as follows: In Section \ref{action}, we outline the Markov-Mukhanov action and derive the background equations. In Section \ref{pert}, we perturb the model to obtain the equation describing the evolution of the density perturbations. In Section \ref{param}, we introduce the model-independent parametrization of Newton's constant and 
in Section \ref{sec:methods}, we discuss how
%provide the plan for incorporating 
the model can be incorporated into the standard data analysis methodology.
Finally, Section \ref{sec:results} presents the findings obtained by fitting the parameterization of $G$ with the data. Throughout the article, we work with the units $c=\hbar=1$.

\section{Theory}
\label{sec:an_fr}
\subsection{Markov-Mukhanov Action}\label{action}

We start with an action where the matter component is non-minimally coupled to gravity via a scalar coupling function that depends on the fluid's density. The multiplicative coupling $\xi(\rho)$ is a function of the energy density $\rho$ and in the GR limit it reduces to $\xi = 1$. The total action  can be written as \citep{Markov:1984nw}
\begin{equation}
    \mathcal{S} = \int \left( \frac{R}{8\pi G_0} + 2 \xi(\rho) \mathcal{L}_m \right)\sqrt{-g}d^4x \ ,
    \label{eq:coupled_action}
\end{equation}
where $\mathcal{L}_m=\rho$ is the matter Lagrangian and
$G_0$ is Newton's gravitational constant at present times. Throughout the manuscript, we adopt the space-like signature $(+---)$ and work with the flat Friedmann-Lemaître-Robertson-Walker (FLRW) metric.

Varying the total action $\mathcal{S}$ with respect to the metric induces a time-varying gravitational constant $G$ and a $\Lambda$-like term on the right-hand side of Einstein's equation as
\begin{equation}
    G_\mu ^ \nu \equiv R_\mu^\nu - \frac{1}{2}\delta_\mu^\nu R = 8 \pi G(\rho) T_\mu^ \nu + \Lambda (\rho) \delta_\mu^ \nu\ .
    \label{eq:mod_Ein}
\end{equation}
Here, the stress-energy tensor $T_\mu^ \nu$ is given by
\begin{equation}
    T_\mu^\nu = (\rho + P) u_\mu u^\nu - P \delta_\mu^\nu,
\end{equation}
where $u_\mu$ is the four-velocity and $P$ is the pressure of the fluid.

The varying gravitational coupling $G(\rho)$ and varying $\Lambda(\rho)$ are related to the gravity-matter coupling function $\xi$ via 
\begin{equation}
        G(\rho) = G_0 \left( \rho\frac{d\xi}{d\rho} + \xi \right), 
        \quad
        \Lambda (\rho) = -8\pi G_0\rho^2 \frac{d\xi}{d\rho}\ .
\label{eq:G_L}
\end{equation}
Notice that we can recover the standard field equations for $\Lambda$CDM cosmology by choosing the scalar coupling function as
\begin{equation}
    \xi_{\Lambda CDM} = \left(1 + \frac{\lambda_0}{\rho}\right)\ ,
    \label{eq:chi_L}
\end{equation}
where $\lambda_0$ is a constant. Then $G$ and $\Lambda$ become both constant with
$G = G_0$ and  $\Lambda = 8\pi G_0 \lambda_0=\Lambda_0$.

The Friedmann equations for the MM model can be written as
\begin{equation}
    \begin{aligned}
        H^2&= \frac{8\pi G(\rho)}{3}\rho + \frac{1}{3}\Lambda(\rho) = \frac{8\pi G_0}{3}  \xi \rho \ , \\
        \frac{\ddot{a}}{a}&= -\frac{4\pi G_0}{3}  \left[ (\rho+3P)\xi + 3\rho \frac{d\xi}{d\rho}(\rho+P ) \right]\ .
    \end{aligned}
    \label{eq:Friedmann}
\end{equation}

Additionally, one can show that the Bianchi identity holds and the total stress-energy tensor (i.e. the r.h.s. of the Eq.~\eqref{eq:mod_Ein}) for the matter component of the universe is conserved,
\begin{align}
    \nabla_\nu \left(8\pi G(\rho)T_\mu^ \nu +\Lambda(\rho) \delta_\mu^ \nu\ \right) = 0. \label{eq:m_conservation}
\end{align}
One can also show that the $ \mu = 0 $ component of the conservation equation reduces to $G(\rho)\nabla_\nu T_0^\nu = 0$, which translates to the continuity equation for the background fluid \citep{Zholdasbek:2024pxi}. In our model, we assume only a single fluid, namely pressure-less dust. Therefore, solving the continuity equation gives $\rho = \rho_\mathrm{m} = \Omega_m \rho_{0}a^{-3}$. Here, $ \rho_\mathrm{m} $ represents the energy density of the pressure-less matter and $\Omega_m$ and $\rho_{0}$ are the density parameter and the total energy density of the universe today, i.e. at $a=1$, respectively. In terms of these parameters, we can write the first Friedmann equation as
\begin{equation}\label{eq:FEDP}
    \frac{H^2}{H_0^2} = \frac{G(a)}{G_0} \frac{\Omega_{\rm m}}{a^3} + \Omega_{\rm \Lambda}(a),
\end{equation}
with $\Omega_{\rm \Lambda}(a) = (\Omega_{\rm m}/3a^2)(d\xi/da)$ and where we have used the monotonic behavior of the density to express $\xi(a)=\xi(\rho(a))$.

As mentioned above, the background quantities are well defined once we adopt a functional form for $G$. We will refer to the model in (\ref{eq:coupled_action}) as GCDM hereafter, since it can effectively replace the cosmological constant with a varying $G$, as we shall show in the following sections.

%%%%%%%%%%%%%%%%%%%%%%%%%%%%%%%%%%%%%%%%%%%%%%

\begin{table*}[!htp]\centering
\caption{A summary of dataset containing independent $f\sigma_8(z)$ measurements from different surveys compiled by \citep{G_parametrized}}\label{tab:dataset}
\fontsize{8}{10}\selectfont
\begin{tabular}{lrrrrrrr}\toprule
Index &Dataset &$z$ &$f\sigma_8(z)$ &Refs. & Fiducial $\Omega_\mathrm{m}$ \\\midrule
1 &6dFGS+SnIa &0.02 &0.428 $\pm$ 0.0465 & \citep{d1} & 0.3 \\
2 &SnIa+IRAS &0.02 &0.398 $\pm$ 0.065 & \citep{d2,d2d3} &  0.3 \\
3 &2MASS &0.02 &0.314 $\pm$ 0.048 & \citep{d3,d2d3}  & 0.266\\
4 &SDSS-veloc &0.1 &0.37 $\pm$ 0.130 &\citep{d4}  & 0.3\\
5 &SDSS-MGS &0.15 &0.490 $\pm$ 0.145 & \citep{d5} & 0.31\\
6 &2dFGRS &0.17 &0.510 $\pm$ 0.060 & \citep{d6}  & 0.3\\
7 &GAMA &0.18 &0.360 $\pm$ 0.090 & \citep{d7d8}  & 0.27\\
8 &GAMA &0.38 &0.440 $\pm$ 0.060 & \citep{d7d8}& 0.27\\
9 &SDSS-LRG-200 &0.25 &0.3512 $\pm$ 0.0583 & \citep{d9d10} & 0.25\\
10 &SDSS-LRG-200 &0.37 &0.4602 $\pm$ 0.0378 & \citep{d9d10}  & 0.25\\
11 &BOSS-LOWZ &0.32 &0.384 $\pm$ 0.095 & \citep{d11} & 0.274\\
12 &SDSS-CMASS &0.59 &0.488 $\pm$ 0.060 & \citep{d12} & 0.307115\\
13 &WiggleZ &0.44 &0.413 $\pm$ 0.080 & \citep{d13d141d15} & 0.27\\
14 &WiggleZ &0.6 &0.390 $\pm$ 0.063 & \citep{d13d141d15} & 0.27\\
15 &WiggleZ &0.73 &0.437 $\pm$ 0.072 & \citep{d13d141d15} & 0.27\\
16 &Vipers PDR-2 &0.6 &0.550 $\pm$ 0.120 & \citep{d16d17} & 0.3\\
17 &Vipers PDR-2 &0.86 &0.400 $\pm$ 0.110 &\citep{d16d17}  & 0.3\\
18 &FastSound &1.4 &0.482 $\pm$ 0.116 &\citep{d18}  & 0.27\\
\bottomrule
\end{tabular}
\end{table*}

%%%%%%%%%%%%%%%%%%%%%%%%%%%%%%%%%%%%%%%%%%%%%%

\subsection{Density Perturbations}\label{pert}

One of the powerful observational probes to discriminate GR from modified theories is the growth of structures as characterized by the matter density contrast $\delta$ \citep{hou2023cosmological}.
%As the next step, w
We now derive the evolution equation for the density contrast $\delta \equiv \delta\rho/\rho$ in the MM prescription, by using linear perturbation theory, which is valid for $\delta \ll 1$. 
Hence, it is convenient to decompose all quantities into 
background (barred) part and perturbations as $X=\bar{X}+\delta X$. The perturbed FLRW metric in the conformal Newtonian gauge with the two scalar metric perturbations $\Psi$ and $\Phi$ reads
\begin{equation}
    g^{\nu \mu}=\bar{g}^{\nu \mu}+\delta g^{\nu \mu}=\frac{1}{a^2}\left(
    \begin{array}{cc}
    1-2 \Psi & 0 \\
    0 & -(1+2 \Phi) \delta^{ij}
    \end{array}\right)\ .
\end{equation}
In a similar manner, we write each term in Eq.~\eqref{eq:mod_Ein} as
\begin{align}
    &G_\mu^\nu=\bar{G}_\mu^\nu+\delta G_\mu^\nu\ , \quad T_\mu^\nu=\bar{T}_\mu^\nu+\delta T_\mu^\nu\ , \quad \notag \\ &G=\bar{G}+\delta G\ , \quad \Lambda=\bar{\Lambda}+\delta \Lambda\ .
    \label{eq:pert_quant}
\end{align}
Substituting Eq.~\eqref{eq:pert_quant} into the field equation
and keeping only zeroth and first-order terms yields 
\begin{equation}
    \bar{G}_\mu^\nu + \delta G_\mu^\nu = 
    8 \pi \bar{G} \bar{T}_\mu^\nu+\bar{\Lambda} \delta_\mu^\nu + 8 \pi \bar{G} \delta T_\mu^\nu + 8 \pi \delta G \bar{T}_\mu^\nu +\delta \Lambda \delta_\mu^\nu\ .
    \label{eq:pert_ein}
\end{equation}
We can now consider LHS and RHS of Eq.~\eqref{eq:pert_ein} separately. 
Since we are looking at large scale perturbations in the matter dominated era, neglecting radiation, we can consider the perturbed energy-momentum to be isotropic. Furthermore,  as can be observed from Eq.~\eqref{eq:pert_ein}, the perturbed $G$ and $\Lambda$ do not lead to any additional anisotropic stresses. Therefore, by neglecting the anisotropic stresses, we can set $\Phi = \Psi$. 
The LHS of equation \eqref{eq:pert_ein} then reads \citep{baumann2022cosmology}
\begin{align}
    G_{00}=&3 \mathcal{H}^2+2 \nabla^2 \Phi-6 \mathcal{H} \Phi^{\prime}\ ,
    \label{eq:00_part}\\
    G_{0 i}=&2 \partial_i \left(  \Phi^{\prime}+\mathcal{H} \Phi  \right)\ , \label{eq:0i_part}\\
    G_{i j}=& -\left(2 \mathcal{H}^{\prime}+\mathcal{H}^2\right) \delta_{i j} 
    + \nonumber \\
    & +\left[2 \Phi^{\prime \prime}+4\left(2 \mathcal{H}^{\prime}+\mathcal{H}^2\right)\Phi + 6 \mathcal{H} \Phi^{\prime}\right] \delta_{i j}\ ,
    \label{eq:ij_part}
\end{align}
where the prime denotes the derivative with respect to conformal time and $\mathcal{H}=a'/a$ is the conformal Hubble parameter. 
Furthermore, if we focus on the subhorizon scale $ k \gg \mathcal{H} $, we can neglect the velocity perturbation, which gives $\Phi^{\prime}+\mathcal{H} \Phi \approx 0$ in Eq.~\eqref{eq:0i_part}. 
By raising the indices in Eq.~\eqref{eq:00_part} and Eq.~\eqref{eq:ij_part} 
while keeping only the first-order terms we get
\begin{eqnarray}
       G_{0}^0 &=& a^{-2}\Big(3 \mathcal{H}^2+2 \nabla^2 \Phi-6 \mathcal{H} ( \Phi^{\prime} + \mathcal{H} \Phi)\Big)  = \notag \\  
       &=& a^{-2}\Big(3 \mathcal{H}^2+2 \nabla^2 \Phi \Big) \ ,\\ 
       G^i_i &=&-3 a^{-2}\left[-\left(2 \mathcal{H}^{\prime}+\mathcal{H}^2\right)+\right. \notag \\ 
       &&+\left. 2\left(\Phi^{\prime \prime}+3 \mathcal{H} \Phi^{\prime}+\left(2 \mathcal{H}^{\prime}+\mathcal{H}^2\right) \Phi\right)\right]\ .
\end{eqnarray}
%=g^{0 0} G_{00} 
%=g^{i \mu} G_{\mu i} =g^{i k} G_{k i}= \\ 

\begin{figure*}[ht]
\centering 
        \includegraphics[width=0.48\textwidth]{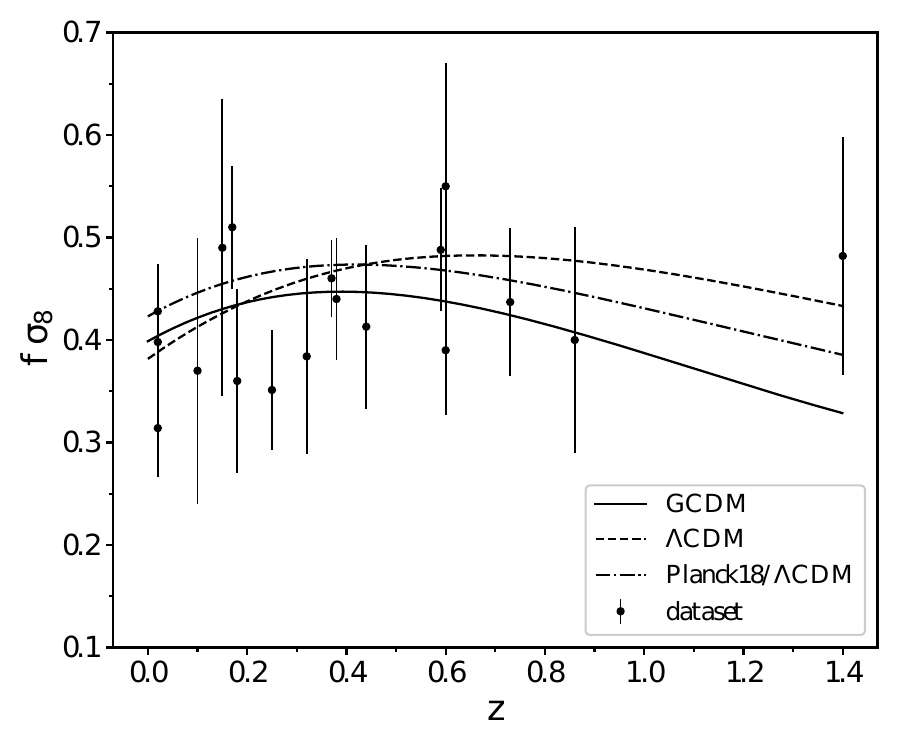}
        \includegraphics[width=0.48\textwidth]{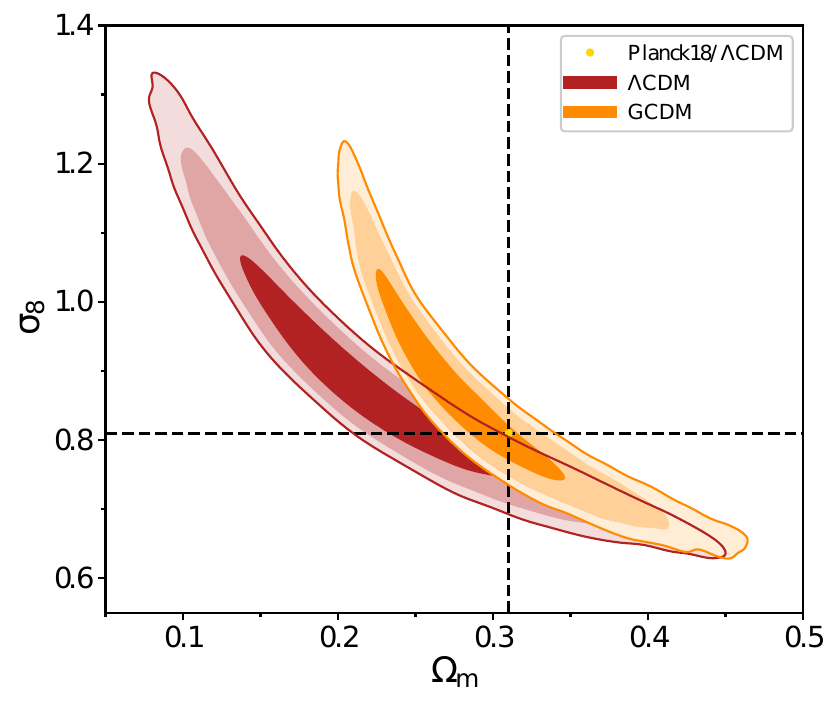}
\caption{Left panel: Vertical lines describe the error bars in the RSD dataset in Table \ref{tab:dataset} . The dashed and dash-dotted lines represent the best fits for $\Lambda$CDM and Planck18/$\Lambda$CDM models, respectively. The solid line represents $G(a)$ parameterization with the best fit $b = 0.24$ corresponding to $n = 3, m = 10$. 
Right panel: Confidence contours obtained from $\Lambda$CDM and $G$CDM models in red and gold, respectively. The yellow dot corresponds to the Planck$18$/$\Lambda$CDM values of $\sigma_8$ and $\Omega_m$.
}
    \label{fig:fs8}
\end{figure*}

On the RHS of Eq.~\eqref{eq:pert_ein}, the first-order terms resulting from matter perturbation are given as
\begin{align}
    \delta T_{0}^0 &= \delta \rho_\mathrm{m}\ ,\\
       \delta T^i_i & = - 3 \delta P_\mathrm{m} \approx 0 \ ,
\end{align}
while dark energy perturbation contributes as
\begin{align}
    \bar{T}_0^0 \delta G+\delta \Lambda \delta_0^0 &= 0\ ,
    \label{eq:dark_pert00}\\
    \bar{T}_j^i \delta G+\delta \Lambda \delta_j^i &= \delta \Lambda 
 \delta_j^i\ .
\end{align}
where $\delta \Lambda = (d\Lambda/d\bar{\rho}_\mathrm{m}) \delta \rho_\mathrm{m}$ and $\delta G = (dG/d\bar{\rho}_\mathrm{m}) \delta \rho_\mathrm{m}$. 
In deriving Eq.~\eqref{eq:dark_pert00}, we used the following relation obtained from Eq.~\eqref{eq:G_L}
\begin{equation}
\rho_\mathrm{m} dG + d\Lambda=0 \ .
\end{equation}

Now, we equate the perturbed Einstein tensor 
and the perturbed total stress-energy tensor.
From the zeroth order terms, 
we obtain Friedmann's first and second equations
as written in Eq.~\eqref{eq:Friedmann}. 
The first-order terms result in
\begin{align}
    \nabla^2 \Phi&=4 \pi \bar{G} a^2 \bar{\rho}_\mathrm{m} \delta\ ,\label{eq:poisson}\\
    \Phi^{\prime \prime}+3 \mathcal{H} \Phi^{\prime}+\left(2 \mathcal{H}^{\prime}+\mathcal{H}^2\right) \Phi&=\Phi^{\prime \prime}+2(\mathcal{H}\Phi)^{\prime}=-\frac{1}{2} a^2 \delta \Lambda\ .
    % = -4 \pi G_0 a^2 \bar{\rho}_\mathrm{m} \frac{d\Lambda}{d\bar{\rho}_\mathrm{m}} \delta
    \label{eq:pot_evol}
\end{align}

In what follows, we will drop the overbar for simplicity.
Recalling that in Fourier mode, we have $\nabla^2 \rightarrow -k^2$ 
where $k$ is the wave number in comoving coordinates, we can 
take time derivatives of Eq.~\eqref{eq:poisson} 
and substitute them into Eq.~\eqref{eq:pot_evol} to obtain

\begin{align}
    &\delta^{''}_k+\mathcal{H} \delta'_k+\left(\mathcal{H}^{\prime}-\mathcal{H}^2 \right) \delta_k=\notag\\ &=- 2 \frac{G'}{G} \delta'_k -\left[\left(1-\frac{k^2}{3 \mathcal{H}^2} \right) \mathcal{H}\frac{G^{\prime}}{G}+\frac{G^{''}}{G}\right] \delta_k\ .
    \label{eq:dens_pert_conformal}
\end{align}

Here, LHS corresponds to the standard GR density perturbation equation while the RHS features contributions characteristic of the model (\ref{eq:coupled_action}) and includes a scale dependent term.

Since we only consider the scales inside Hubble radius ($k>>\mathcal{H}$) and expect $G'\ll 1$, 
all the terms on the RHS of Eq.~\eqref{eq:dens_pert_conformal} become negligible except for the component involving scale dependence, resulting in
\begin{equation}
    \delta^{''}_k+\mathcal{H} \delta'_k+\left[\mathcal{H}'-\mathcal{H}^2 - \frac{k^2}{3 \mathcal{H}} \frac{G'}{G} \right] \delta_k= 0\ .
\end{equation}
Using the second Friedmann equation in comoving coordinates
and converting derivatives from conformal time to scale factor $a$, 
we obtain the main equation of our analysis, 
\begin{align} 
& a^2 \frac{d^2\delta_k}{da^2} 
 + \left(3 + \frac{d\ln{H}}{d\ln{a}}\right) a \frac{d\delta_k}{da} + \notag\\
& - \left(\frac{3}{2} \frac{G}{G_0} \frac{H_0^2\Omega_\mathrm{m}}{a^3 H^2}  
+ \frac{1}{3}\frac{k^2}{ a^2 H^2} \frac{d\ln{G}}{d\ln{a}} \right)\delta_k = 0\ .
\label{eq:dens_pert_a}
\end{align}
In contrast to the $\Lambda$CDM model, here we have an extra term arising because of the variable nature of $G$, which will affect the growth history of the universe throughout its evolution \footnote{To be precise, the effects of a modified theory would also enter Eq.~\eqref{eq:dens_pert_a} via the Hubble parameter $H(a)$.}. Therefore any modification of the standard $\Lambda$CDM scenario would contribute towards the modified growth rate of the density perturbations.

\subsection{Parameterizing $G(\rho)$}\label{param}

We shall now propose a functional form for $G$ to parameterize deviations from GR close to the current epoch.  
In a matter dominated universe, the matter density scales as $\rho_\mathrm{m} \sim a^{-3}$, and by recalling Eq.~\eqref{eq:G_L} we find
\begin{equation}
    G = G_0\left(-\frac{a}{3}\frac{d\xi}{da} + \xi \right).
    \label{eq:G_chi_a}
\end{equation}
In our model, we assume that at very early times i.e. $a \sim 0$, the universe was purely matter dominated for a sufficiently long period, meaning that we must impose $\xi \sim 1$. This translates to two conditions,
as $ a \rightarrow 0$, $G \simeq G_0$ and $dG/da \simeq 0$. On the other hand, we would also like to recover $G \simeq G_0$ today, i.e. at $a=1$. These conditions impose substantial restrictions on the parameterized form of $G(a)$. 

The lunar laser experiment further limits the running of the gravitational constant at present times to $dG/da|_{(a=1)} \approx 0.001$ \citep{lunar_laser}. Meanwhile, Big Bang Nucleosynthesis (BBN) requires 
$G = 1.09 \pm 0.2$ 
to remain consistent with the results of 
 \citep{Bambi_2005}.
We propose the following simple parameterization,
which is designed to accommodate all these constraints
\begin{equation}
    G/G_0 = 1 + b a^n (1-a)^{2m} ,
    \label{G_param}
\end{equation}
where $b$ is the free parameter, $n$ and $m$ takes only integer values. The $\Lambda$CDM limit of Eq.~\eqref{eq:chi_L} can be recovered by setting $b=0$ and integrating Eq.~\eqref{eq:G_chi_a}.

In the next section, we shall use $f\sigma_8$, a quantity constructed out of the growth rate $f$, to explore the growth history of the model. We shall then compare the parameterized model with observational data in order to test its ability to ease the $\sigma_8$ tension.

\begin{figure*}[ht]
\centering 
        \includegraphics[width=0.48\textwidth]{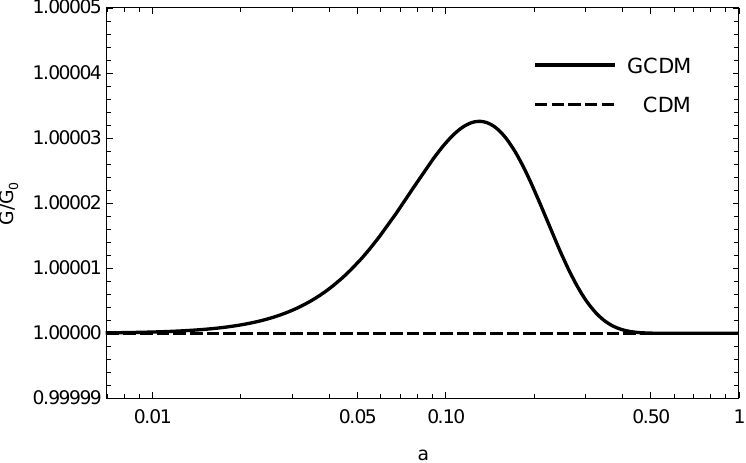} 
        \includegraphics[width=0.45\textwidth]{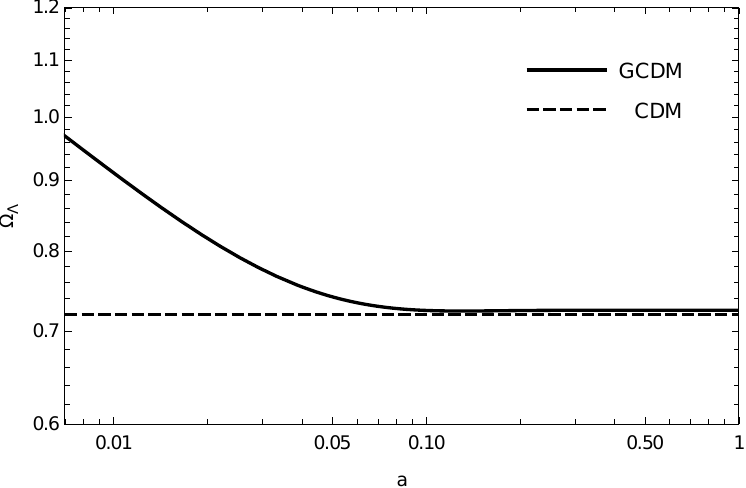} 
        \caption{Left panel: The variation of $G(a)$ with respect to $a$ for GCDM (solid line) and $\Lambda$CDM (dashed line). Right panel: The variation of $\Omega_{\Lambda}(a) = (\Omega_{\rm m}/3a^2)(d\xi/da)$ with respect to $a$ (solid line) as compared to a constant $\Lambda$ (dashed line). Here, we use the bestfit parameters $\Omega_{\rm m} = 0.28$, $b = 0.24$, $n=3$ and $m=10$.
}
    \label{fig:G-Lambda}
\end{figure*}

\section{Numerical Analysis}\label{nummeth}
\subsection{Method}
\label{sec:methods}

The growth rate of density fluctuations $f$ has been measured by several cosmological surveys in the redshift range $z \in (0.02 , 1.4)$. It is defined as 
\begin{equation}
    f(a) = \frac{d\delta_k(a)}{d\ln a}.
\end{equation}
The $\sigma_8$ parameter is commonly used to illustrate the lack of power of Planck18/$\Lambda$CDM on small scales. However, from the perspective of Redshift Space Distortion (RSD) measurements, $f\sigma_8$ is regarded as more reliable and independent of bias than $\sigma_8$ alone \citep{Song_2009}. 
We construct this new parameter from the solution of Eq. \eqref{eq:dens_pert_a} as 
\begin{equation}
    f\sigma_8 = \sigma_8 \frac{d\delta_k}{d\ln a} .
\end{equation}
Next, we leverage Bayesian analysis to test the proposed model with a set of independent $f\sigma_8(z)$ measurements from different surveys (see Table \ref{tab:dataset}). Our aim is to find the best fit values for $b$ and $\sigma_8$ by minimizing
\begin{equation}
    \chi^2= VC^{-1}V^T\ ,
\end{equation}
where $C^{-1}$ is the inverse covariance matrix of the data and $V$ is the raw vector, in which $i$-th column is given by
\begin{equation}
    V_i= f\sigma_{8}(z_i)-\textrm{ratio}(z_i)f\sigma_8(z_i,b,\sigma_8)\ , \quad i=1,\ldots,N,
\end{equation}
where $N$ is the total number of measurements in the dataset, $f\sigma_{8}(z_i)$ corresponds to $i$-th value in the dataset for $z_i$, $f\sigma_8(z_i,b,\sigma_8)$ is the predicted value given a set of parameter values at $z_i$. As observed in Table \ref{tab:dataset}, various collaborations implemented different fiducial cosmologies to convert redshifts to distances. To account for this model dependence, the term $\textrm{ratio}(z_i)$ is introduced as the ratio between the products of the Hubble parameter $H(z)$ and the angular diameter distance $d_A(z)$ of the model at hand to that of the fiducial cosmology \citep{Macaulay_2013}
\begin{equation}
    \textrm{ratio}(z) = \frac{H(z)d_A(z)}{H^{fid}(z)d^{fid}_A(z)}\ .
\end{equation}

It is assumed that all data are uncorrelated, except WiggleZ data, for which the covariance matrix is given by \citep{Blake_2012}
\begin{equation}
    C^{\textrm{WiggleZ}} = 10^{-3}\left(
    \begin{array}{ccc}
    6.400 & 2.570  & 0.000 \\ 
    2.570 & 3.969 & 2.540 \\ 
    0.000 & 2.540 & 5.184
    \end{array}\right)\ .
\end{equation}
Thus, we have a $N\times N$ covariance matrix with the $3\times 3$ submatrix $C^{\textrm{WiggleZ}}$ at the position of the WiggleZ data.

We use open-source sampler \texttt{emcee} \citep{emcee} along with \texttt{GetDist} \citep{getdist} libraries to obtain the best-fit cosmological parameters. The Python implementation of our analysis can be found on the GitHub\footnote{\url{https://github.com/Tilek-Zhumabek/GCDM}} repository.

\subsection{Results}
\label{sec:results}

In this section, we present our findings by fitting the predicted values of $f\sigma_8$ with the RSD data to estimate $\sigma_8$. 
The scale dependence in Eq.~\eqref{eq:dens_pert_a} is handled by fixing $k = 300 H_0 \sim 0.1 h$ Mpc$^{-1}$. 
This choice is motivated by the fact 
that nonlinear effects become significant starting from scales 
smaller than $k = 0.2 h$ Mpc$^{-1}$, 
while we still have limited measurement accuracy on scales larger than $k = 0.01 h$ Mpc$^{-1}$\citep{Zhang_2012, Arjona_2019}. 

Fig.~\ref{fig:fs8} presents the best fit of $\Lambda$CDM and GCDM models to the $f\sigma_8$ dataset of Table \ref{tab:dataset}.
The solid line is the $G$ parameterization 
($n = 3, m=10$) with the best fit values 
$(b = 0.24^{+0.089}_{-0.097}, \Omega_\mathrm{m} = 0.28^{+0.092}_{-0.073}, \sigma_8 = 0.87^{+0.20}_{-0.19}, \chi^2 = 14.2)$,
the dash-dotted line corresponds to Planck18/$\Lambda$CDM model $(\Omega_\mathrm{m} = 0.31^{+0.0074}_{-0.0074}, \sigma_8 = 0.81^{+0.0061}_{-0.0061},\chi^2 = 17.6)$,
and the dashed line represents the best fit parameter values $(\Omega_\mathrm{m} = 0.22^{+0.12}_{-0.11}, \sigma_8 = 0.89^{+0.24}_{-0.20}, \chi^2 = 12.3)$ 
of $\Lambda$CDM.  

Note that the scale-dependent part in Eq.~\eqref{eq:dens_pert_a} plays a pivotal role in the structure growth, which is not present in the standard $\Lambda$CDM scenario.
%GR neglects due to the gravitational constant $G' = 0$. 
However, a slight deviation from GR results in a meaningful contribution from that term which can alleviate the $\sigma_8$ tension. The overall effect can bring the tension level 
from $3 \sigma$ to within $1 \sigma$ as illustrated on the right panel of Fig.~\ref{fig:fs8}. 

In Fig.~\ref{fig:G-Lambda}, we show the behaviour of the parameterized $G$ and the induced dark energy density $\Omega_\Lambda (a)$ with the the best fit values of the parameters. The left panel shows the running nature of $G$ with $b = 0.24$, $n=3$ and $m=10$ (solid line) compared to the constant value for GR (dashed line). We can see that the model requires a slightly higher value of $G$ around $a \sim 0.01 - 0.5$ to alleviate the $\sigma_8$ tension. This running nature of $G$ induces a dark energy component in the universe. This can be seen from the Friedmann equation \eqref{eq:FEDP} where the induced dark energy component is given by
\begin{equation}
    \Omega_{\rm \Lambda}(a) = \frac{\Omega_{\rm m}}{3a^2}\frac{d\xi}{da}.
\end{equation} 

We plot this dark energy density in the right panel of Fig.~\ref{fig:G-Lambda} with $b = 0.24$, $n=3$ and $m=10$ (solid line). The dashed line represents the constant value for $\Lambda$CDM. To alleviate the $\sigma_8$ tension, the model requires a slightly increased dark energy density in the early universe. The combination of the running natures of both G(a) and $\Omega_\Lambda(a)$ helps ease the $\sigma_8$ tension by
%. Basically, the running nature 
activating the scale-dependent part of the growth equation. This initially enhances the source term in Eq.~\eqref{eq:dens_pert_a} and subsequently diminishes it, corresponding to $G'(a)$ being positive initially and then becoming negative. 

%{\bf (DM: Should we comment on the potential impact on Hubble tension as opposed to other models?)}

%In Fig.~\ref{fig:G}, we show the behaviour of our parameterization  with the best fit $b = 0.24$ value and for various $n$ and $m$.
%Notably, even a fifth-order difference in $G(a)$ (the best fit solid black line) from unity can activate the scale-dependent part of the growth equation. 
%This initially enhances the source term in Eq.~\ref{eq:dens_pert_a} 
%but subsequently diminishes it, 
%corresponding to $G'(a)$ being positive initially 
%and then becoming negative. 

% It is important to highlight that in our best fit $G(a)$ parameterization ($b=0.24, n= 3, m=10$), it is crucial to keep $n = 3$ while the choice $m = 10$ is not unique in reducing the tension. 
% Alternative $m$ values close to $m=10$ can also effectively alleviate the tension since $m$ has a very small effect on $G$ compared to $n$ as can be seen in Fig.~\ref{fig:G}.
% {\bf (HC: If we want to show Fig.~\ref{fig:G} in the final draft, then it is better to make the y-axis dimensionless. I would leave it upto Tilek. Try not to use coloured lines. Use solid, dashed and dotted. Accordingly change the above para. )}
% \tilek{For me, I am fine to get rid of this paragraph altogether along with Fig.3.}

\section{Conclusion}
\label{sec:conclusion}

Structure growth is a powerful tool to test deviations from GR via observations, as any modified gravity approach has a distinct signature in the evolution of perturbed densities. 
In this work, we examined the implication of one of the simplest modifications to GR, namely a running gravitational coupling, on the low redshift universe.
We defined the background quantities in the modified scenario and conducted a full perturbation analysis for an arbitrary
%Subsequently, we introduced a 
parameterization of $G(a)$ close to today
%that mimics a Taylor expansion, 
to fit the model parameters with the RSD dataset. 

Our findings indicate that at the scale $k=0.1 h$ Mpc$^{-1}$, GCDM effectively reduces the tension level to within $1 \sigma$. Remarkably, achieving this result requires a minuscule deviation in $G(a)$, equivalent to a fifth-order difference, from Newton's gravitational constant. This running nature of $G(a)$ induces a dark energy component in the universe, which settles to the $\Lambda$CDM value at late times. 

It is well known that the $\sigma_8$ tension is correlated to the $H_0$ tension. It can happen that models that alleviate the $H_0$ tension, can aggravate the $\sigma_8$ tension \citep{DiValentino:2020vvd}.
Therefore, it is necessary to perform a conjoined analysis of the model in the context of $\sigma_8$ and $H_0$ tension to constrain the ranges of the model's parameter for which the resulting evolution history of the universe remain within accepted experimental bounds over the given period \citep{Linder:2016xer}.
%because conjoint history can deviate significantly at intermediate scales \citep{Linder:2016xer}. 
For this purpose, in future analysis we aim to compare our model's predictions to other cosmological tests at different epochs such as Baryon Acoustic Oscillations, CMB, galaxy power spectrum shape and void measurements etc.

\section*{Acknowledgments}
TZ acknowledges the Galileo Galilei Institute for Theoretical Physics for the hospitality and the INFN for partial support during the completion of this work. DM and HC acknowledge support from Nazarbayev University Faculty Development Competitive Research Grant No. 11022021FD2926. Also, the authors would like to acknowledge Prof. Ernazar Abdikamalov for his constructive feedback and valuable suggestions throughout the writing of this paper.

\bibliographystyle{apsrev}
\bibstyle{apsrev}

\bibliography{templateArxiv}

\end{document}